\documentclass[journal,12pt,draftclsnofoot,onecolumn]{IEEEtran}
\IEEEoverridecommandlockouts
\ifCLASSINFOpdf
\else
\fi
\usepackage{graphicx}
\usepackage{latexsym,bm}
\usepackage{amsmath}
\usepackage{amssymb}
\usepackage{amsthm}
\usepackage{epstopdf}
\usepackage{cite}
\usepackage{mathrsfs}
\usepackage{multirow}
\usepackage[margin=1.0in]{geometry}
\usepackage{booktabs}
\usepackage{array}
\usepackage{caption}
\usepackage{pgfplots}
\usepackage{algorithm}
\usepackage{pbox}
\usepackage{epsfig}
\usepackage{algorithmic}
\usepackage{subcaption}
\pgfplotsset{compat=newest} 
\pgfplotsset{plot coordinates/math parser=false}
\newcolumntype{M}{ >{\centering\arraybackslash} m{8cm} }
\newcolumntype{L}{ >{\centering\arraybackslash} m{7.5cm} }
\newcolumntype{K}{ >{\centering\arraybackslash} m{\textwidth} }

\usepackage{lipsum}

\newcommand\blfootnote[1]{%
  \begingroup
  \renewcommand\thefootnote{}\footnote{#1}%
  \addtocounter{footnote}{-1}%
  \endgroup
}

\begin{document}
\begin{titlepage}
\begin{center}
\vspace*{-2\baselineskip}
\begin{minipage}[l]{7cm}
\flushleft
\includegraphics[width=2 in]{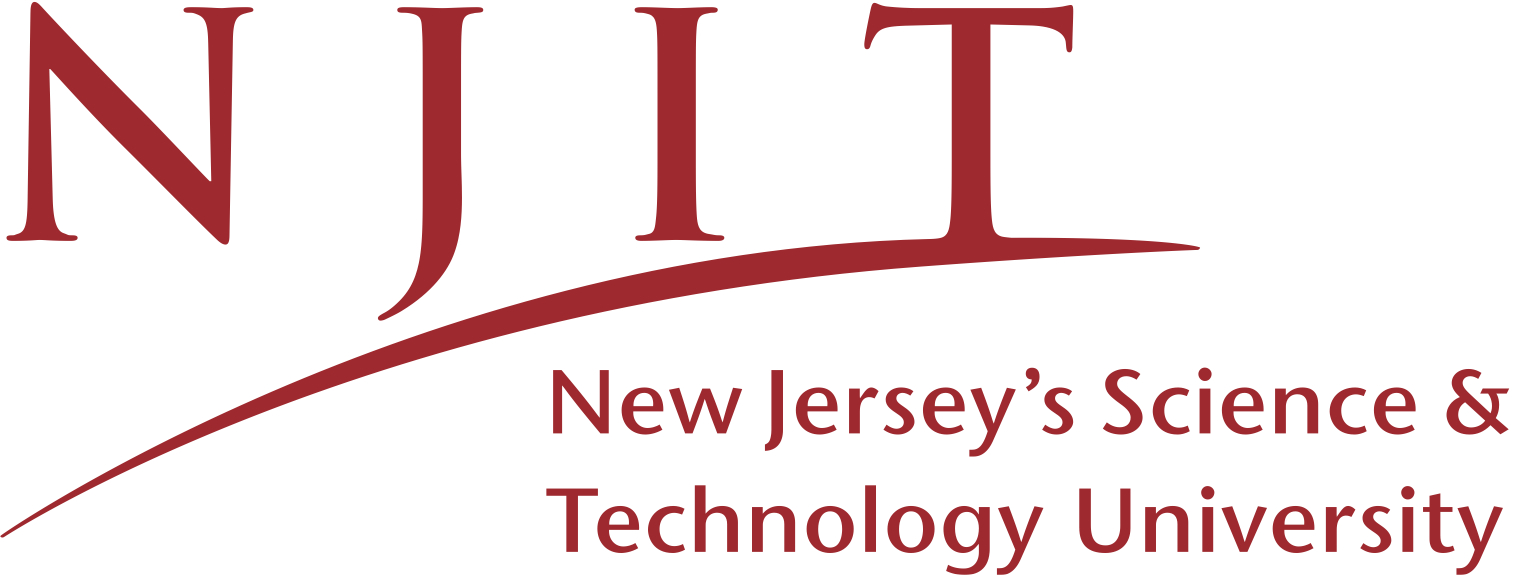}
\end{minipage}
\hfill
\begin{minipage}[r]{7cm}
\flushright
\includegraphics[width=1 in]{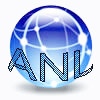}%
\end{minipage}

\vfill

\textsc{\LARGE RF Energy Harvesting Enabled \\[12pt]
Power Sharing in Relay Networks }\\

\vfill
\textsc{\LARGE Xueqing Huang\\[12pt]
\LARGE NIRWAN ANSARI}\\
\vfill
\textsc{\LARGE TR-ANL-2014-008\\[12pt]
\LARGE September 29, 2014}\\[1.5cm]
\vfill
{ADVANCED NETWORKING LABORATORY\\
 DEPARTMENT OF ELECTRICAL AND COMPUTER ENGINEERING\\
 NEW JERSY INSTITUTE OF TECHNOLOGY}
\end{center}
\end{titlepage}

\begin{abstract}
Through simultaneous energy and information transfer, radio frequency (RF) energy harvesting (EH) reduces the energy consumption of the wireless networks. It also provides a new approach for the wireless devices to share each other's energy storage, without relying on the power grid or traffic offloading. In this paper, we study RF energy harvesting enabled power balancing within the decode-and-forward (DF) relaying-enhanced cooperative wireless system. An optimal power allocation policy is proposed for the scenario where both source and relay nodes can draw power from the radio frequency signals transmitted by each other. To maximize the overall throughput while meeting the energy constraints imposed by the RF sources, an optimization problem is formulated and solved. Based on different harvesting efficiency and channel condition, closed form solutions for optimal joint source and relay power allocation are derived.
\end{abstract}



\blfootnote{This work was supported in part by NSF under grant no. CNS-1320468.}

\section{Introduction}
\emph{Green powered wireless network} is of great social, environmental, and economic potential because wireless access networks are among the major energy guzzlers of the telecommunications infrastructure, and their current rate of power consumption is escalating because of the explosive surge of mobile data traffic 
\cite{6472203} - \cite{Han:2012:ICE}. To assuage the dependence on the traditional unsustainable energy, the concept of energy harvesting (EH) has been proposed as a key enabling technology. 

From wind, solar, biomass, hydro, geothermal, tides, and even radio frequency signals \cite{RF}, EH is capable of generating electricity or other energy form, which is renewable and more environmentally friendly than that derived from fossil fuels \cite{RFHARVESTOR}. If the green energy source is ample and stable in the sense of availability, the wireless network can be powered by the harnessed free energy permanently, without requiring external power cables or periodic battery replacements.

To guarantee a certain level of stability in energy provisioning, \emph{hybrid powered devices} often require a backup non-renewable energy source for the energy harvesting generators \cite{Han:2012:ICE}, and \emph{passively powered devices} normally draw multiple green energy sources in a complementary manner \cite{RF}. However, the energy still cannot be consumed before it is harvested. As compared with stable on-grid energy, \emph{opportunistic energy harvesting} results in fluctuating power budget, namely, \emph{energy causality constraint} (EC-constraint). The EC-constraint mandates that, the total consumed energy should always be no greater than the total harvested energy, which maybe further limited by the finite battery capacity \cite{5522465}, \cite{energyCausality}.

To maximize the system performance while not violating the EC-constraint for the architecture with separated energy harvester and information transmitter, Ho and Zhang \cite{6202352} considered the point-to-point wireless system with the energy harvesting transmitter. Optimal energy allocation algorithms are developed to maximize the throughput over a finite time horizon. Similarly, the throughput by a deadline is maximized and the transmission completion time of the communication session is minimized \cite{5441354}, \cite{5992841}. Moreover, the works in \cite{6363767} and \cite{6381384} explored the joint source and relay power allocation over time to maximize the throughput of the three node decode-and-forward (DF) relay system, in which both the source and relay nodes transmit with power drawn from independent energy-harvesting sources.

Among all of the green energy sources, radio frequency (RF) energy harvesting provides a new approach for short distance energy sharing in lieu of traffic offloading \cite{aaaa} or traditional power grid. To balance the power consumption of the wireless network, mobile charging systems can deploy mobile vehicles/robots, which carry high volume batteries \cite{6502505}, to serve as back up mobile power storage and periodically deliver energy to wireless devices with insufficient energy supply.

Another characteristic of RF energy harvesting is the provisioning of simultaneous transfer of wireless information and power \cite{6489506}. The \emph{separated} data decoder and energy harvester can receive data and harness energy from the received RF signals. The \emph{co-located} data and energy reception components can either split the common received signals (\emph{power splitting}), or perform the above mentioned two processes sequentially (\emph{time switching}) \cite{6552840}. 

To capitalize on wireless energy sharing and simultaneous data and energy transmission, we analyze the half-duplex relay system, where the source node (SN) and relay node (RN) can harvest energy from each other. In particular, the relay node can simultaneously harvest energy and receive data from the signals transmitted by the source node in the first time slot, and source node can harvest energy from the forwarding signals transmitted by the relay node in the second time slot. Depending on the residual energy levels, the power sharing within the system can be facilitated by adjusting transmission power of both SN and RN.

The rest of the paper is organized as follows. After introducing the system with radio frequency energy harvesting enabled relay system in Section II, we analyze joint energy management policies for both the source and relay nodes in Section III. We derive the optimal power allocation, which maximizes the system throughput, in Section IV. Then, numerical results are presented in Section V. Conclusions are given in Section VI.

\section{System Scenario with RF-EH enabled Relay and Source Nodes}
\begin{figure}
\centering
\includegraphics[width=2.3 in]{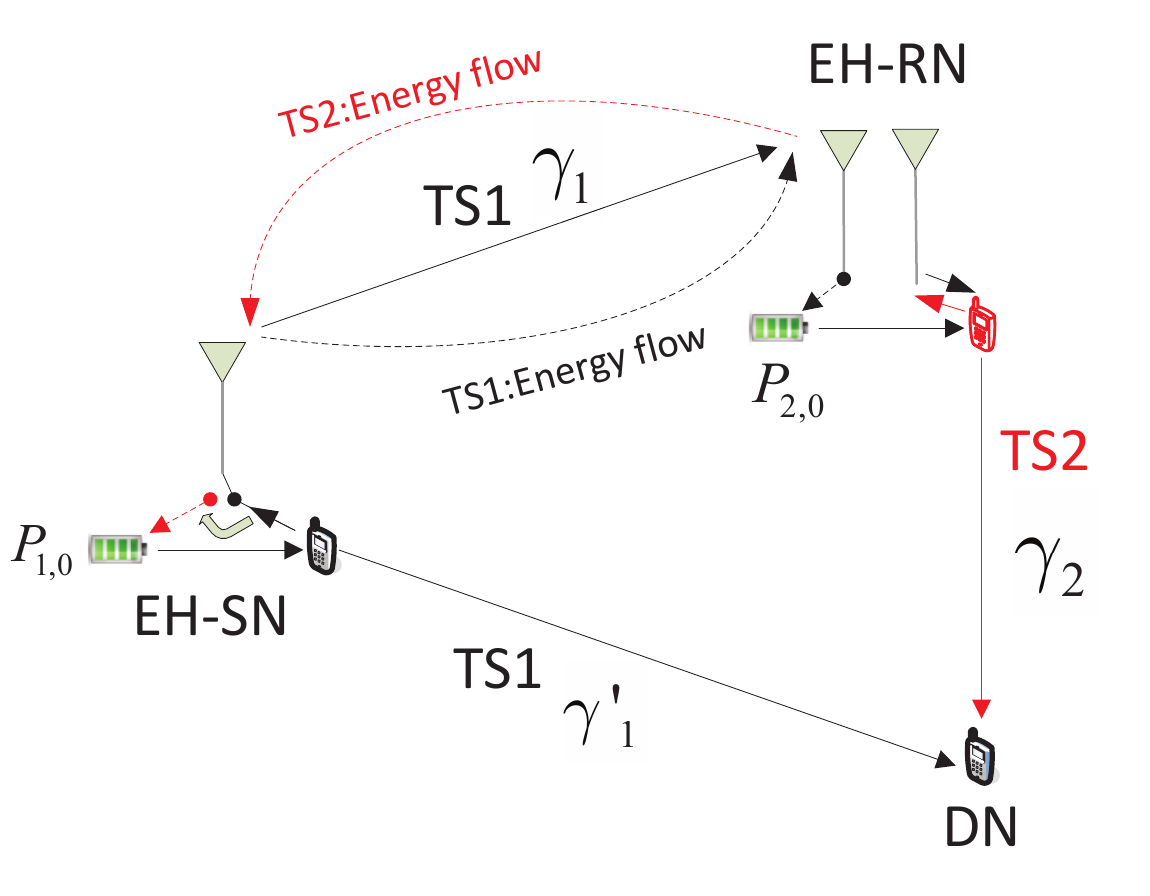}
\vspace{-1em}
\caption{RF-EH enabled DF-relay system.}
\label{scenario}
\vspace{-1.5em}
\end{figure}

Consider the Shannon capacity of the half duplex relay system measured over $N$ phases, where $N$ can be the delay requirements of data traffic, and each phase consists of two consecutive time slots (TSs). As illustrated in Fig. \ref{scenario}, in each odd TS, SN transmits data to the relay node, while in the even TS, RN forwards the signal received in the previous TS. The amount of the green/brown energy already acquired by SN and RN are $P_{1,0}$ and $P_{2,0}$, respectively. The energy harvested from the RF signals can be used to facilitate future data transmission. 

The total bandwidth occupied by the system is $B$. For the sake of convenience, we assume the constant channel power gains across $N$ phases \cite{6381384}, where $h_{i}$ is the channel gain of the SN-RN link ($i=1$) and the RN-DN link ($i=2$). ${\gamma_{i}} = {{{{\left| {{h_{i}}} \right|}^2}}}/({{{N_0}B}})$ denotes the corresponding normalized signal-noise-ratio (SNR) associated with the channel between SN and RN ($i=1$) as well as that associated with the channel between RN and DN ($i=2$). ${N_0}B$ represents the power of additive white Gaussian noise. Without loss of generality, for now, we assume no direct link exists between SN and DN, i.e., the corresponding SNR $\gamma_1'=0$. 

The goal is to design the optimal power allocation $P_{i,j}$, $i = \{1,2\},j\in\{1,\cdots,N\}$ such that the overall system throughput cross $N$ phases is maximized.

\begin{equation}
\label{initial}
\begin{array}{l}
{C^*} = \mathop {\max }\limits_{\{P_{i,j}\}} C = \frac{B}{2}\sum\limits_{j = 1}^N {\mathop {\min }\limits_{i = 1,2} \{ \log (1 + {P_{i,j}}{\gamma _i})} \} \\
\begin{array}{*{20}{l}}
{s.t.}&EC_{1,j}:&\sum\limits_{k = 1}^j {{P_{1,k}}}  \le {P_{1,0}} + \beta \sum\limits_{k = 1}^{j - 1} {{P_{2,k}}}\\
&EC_{2,j}:&\sum\limits_{k = 1}^j {{P_{2,k}}}  \le {P_{2,0}} + \beta \sum\limits_{k = 1}^{j} {{P_{1,k}}}\\
&{NC}:&{P_{i,j}} \ge 0,i\in\{1,2\},j\in\{1,\cdots,N\}
\end{array}
\end{array}
\end{equation}
where $EC_{i,j}$, $j \in \{ 1, \cdots ,N\}$ is the energy causality constraint of the $j$-th phase in SN ($i=1$) and RN ($i=2$), respectively. Constraint $NC$, represents the non-negative power allocation. ${B}/{2}$ is attributed to the half-duplex of the relay channel. $\beta{P_{i,j}}$ is the amount of power harvested in phase $j$ by SN ($i=2$) and RN ($i=1$), respectively, and $\beta={{\eta}} {{\left| {{h_1}} \right|}^2}$ with $\eta$ denoting the energy harvesting efficiency factor \cite{6702854}.

Note that the difference between $EC_{1,j}$ and $EC_{2,j}$ implies that RN can use power harvested in phase $j$, while SN can only use power harvested before phase $j$. This is because the energy consumption process of RN, i.e., data forwarding, is one TS after the simultaneous energy harvesting and data reception processes.

\section{Power Allocation Analysis}\label{section2}
Since the throughput in phase $j$ is determined by $\mathop{\min }\limits_{i=1,2}\{{P_{i,j}}{\gamma_i}\}$, we can divide $P_{i,j}$ into the \emph{power supplement} part $\alpha_{i,j}$ and \emph{data transmission} part $p_{i,j}$.
\[\left\{\begin{array}{l}
P_{i,j}=p_{i,j}+\alpha_{i,j},i\in\{1,2\}\\
p_{1,j}\gamma=p_{2,j}\text{, }\gamma={\gamma_1}/{\gamma_2}
\end{array}\right.\]
where 1) if ${P_{1,j}}{\gamma _1}\ge{P_{2,j}}{\gamma _2}$, then, ${p_{2,j}}=P_{2,j}$, ${a_{2,j}}=0$, and $p_{1,j}={P_{2,j}}/{\gamma}$ is used to transmit data that will be forwarded by RN; ${\alpha_{1,j}}={P_{1,j}}-p_{1,j}$ is used to increase the energy storage in RN. 2) If ${P_{1,j}}{\gamma_1}<{P_{2,j}}{\gamma _2}$, then, ${p_{1,j}}=P_{1,j}$, ${a_{1,j}}=0$, and $p_{2,j}={P_{1,j}}{\gamma}$ is used to forward the data, and ${\alpha_{2,j}}={P_{2,j}}-p_{2,j}$ is used to increase the energy storage in SN.

\subsection{$\beta\ge\gamma$}
In this case, regardless of how much power SN uses in each phase to transmit data, RN can harvest $P_{1,j}\beta$ amount of power from the received signal, which is greater than $P_{1,j}\gamma$, i.e., the amount of power needed to forward data in the $j$-th phase. So, as compared with SN, the amount of power in RN is always sufficient, and there is no need to provide power supplement to RN, i.e., $a_{1,j}=0$, $j\in\{1,\cdots,N\}$. Consequently, the system throughput is determined by $P_{1,j}$, and RN will adopt the \emph{fully cooperative strategy}: transmit all of its residual power in each phase to increase the power storage in SN. The optimization problem in Eq. (\ref{initial}) is simplified as Eq. (\ref{obj_fin_relay}), and the solution will be discussed in Section IV-B.
\begin{equation}
\label{obj_fin_relay}
\begin{array}{l}
\mathop {\max }\limits_{\{ {P_{1,j}}\} }\sum\limits_{j = 1}^N {\log (1 + {P_{1,j}}{\gamma _1})} \\
\begin{array}{*{20}{l}}
{s.t.}&{P_{2,1}}=P_{1,1}\beta+P_{2,0}\\
{}&{P_{2,j}}=P_{1,j}\beta,j\in\{2,\cdots,N\}\\
EC_{1,1}:&{{P_{1,1}} \le {P_{1,0}}}\\
EC_{1,j}:&{\sum\limits_{k = 1}^j {{P_{1,k}}}  \le {P_{1,0}}+{P_{2,0}}\beta + \beta^2 \sum\limits_{k = 1}^{j - 1} {{P_{1,k}}}}\\
\end{array}
\end{array}
\end{equation}
where $EC_{1,j}$, $j\in\{2,\cdots,N\}$ is the EC-constraint of SN with the fully cooperative RN in the $j$-th phase. $EC_{1,N}$ is satisfied with equality. 

\subsection{$\beta<\gamma$}
In this case, although RN can directly use the energy harvested in the first TS to forward data, $(\gamma-\beta)P_{1,j}$ amount of the \emph{residual power} still will be consumed in the second TS, when $\alpha_{1,j}=0$. The residual power means the power that is already in the battery of SN ($i=1$) or RN ($i=2$) at the beginning of phase $j$, which is denoted as $\overline{P_{i,j}}$ for the rest of the paper. Therefore, in the $j$-th phase, if RN provides power supplement while not receiving power supplement, $\alpha_{2,j}$ must come from $\overline{P_{2,j}}$. Since $\overline{P_{2,j}}$ can be consumed in phase $j-1$, RN can have the following equivalent power supplement allocation.
\begin{equation}\label{ite}
\alpha^*_{2,j-1}=\alpha_{2,j-1}+\alpha_{2,j}\text{, }\alpha^*_{2,j}=0
\end{equation}

Although power supplement enables SN and RN to share each other's power, in case one of their residual energy is insufficient, $\beta$ (normally less than 1) means power loss of the system will increase with power supplement. Therefore, if $\alpha_{1,j}\alpha_{2,j}>0$, aggregating power supplement to SN or RN will save both of them some power.
\begin{equation}
\label{compare}
\alpha^*_{1,j}=\{\alpha_{1,j}-\alpha_{2,j}\}^+\text{, }\alpha^*_{2,j}=\{\alpha_{2,j}-\alpha_{1,j}\}^+\\
\end{equation}
where $\{\bullet\}^+=\max\{0,\bullet\}$.

\emph{Proposition 1:} With $\beta<\gamma$, any power supplements provided by RN can be aggregated to the first phase.
\[\left\{\begin{array}{*{20}{l}}
\alpha_{2,1}=\alpha_2\ge0\\
\alpha_{2,j}=0, j\in\{2,\cdots,N\}
\end{array}\right.\]
where $\alpha_2$ is the aggregated \emph{power supplement} transmitted by RN in the first phase.
\begin{proof}
If $\exists j\in\{2,\cdots,N\}$ such that $\alpha_{2,j}>0$, then according to Eqs. (\ref{ite}) and (\ref{compare}), $\alpha_{2,j}$ can be aggregated to either $\alpha_{2,j-1}$ or $\alpha_{1,j}$, such that $\alpha^*_{2,j}=0$.
\end{proof}
\emph{Remark 1:} If $\alpha_2>0$, then $\alpha_{1,1},\alpha_{1,2}=0$.

With different system parameter $\beta$ and $\gamma$, we find the optimal power allocation with aggregated $\alpha_2$, i.e., $P_{1,j}\ge{P_{2,j}}/\gamma$, $j\in\{2,\cdots,N\}$, in the next section.

\section{Optimal Power Allocation with $\beta<\gamma$}\label{section3}
\subsection{$\beta\gamma\ge1$}
With $\beta\gamma\ge 1$, regardless of how much power SN uses to transmit data, if RN has sufficient energy to match $P_{1,j}$, SN will harvest $P_{1,j}\beta\gamma$ amount of power for future data transmission, which is greater than the energy spent in phase $j$. So, SN prefers to adopt the \emph{fully greedy strategy}: transmit all of its residual power in each phase to increase the energy storage of RN. However, when adopting the fully greedy strategy, the residual energy of SN will only be the energy harvested from phase $j-1$, $\overline{P_{1,j}}=\beta{P_{2,j-1}}$, $j\in\{2,\cdots,N\}$. We check whether $\overline{P_{1,j}}$ is sufficient for the data transmission in the $j$-th phase by presenting the following proposition.

\emph{Proposition 2:} There exists an optimal power allocation which satisfies the following inequality. 
\[P_{2,j}\beta\gamma\ge{P_{2,j+1}}, j\in\{1,\cdots,N-1\}\] 
\begin{proof}
Suppose $\exists j\in\{1,\cdots,N-1\}$ such that $P_{2,j}\beta\gamma<{P_{2,j+1}}$. Since $\alpha_{2,j+1}=0$ (\emph{Proposition 1}), there exists
\begin{equation}\label{useful}
p_{1,j+1}>\beta{P_{2,j}}\ge{p_{1,j}}\text{, }p_{2,j+1}>{P_{2,j}\beta\gamma}\ge{p_{2,j}}
\end{equation}
Then, decreasing $p_{i,j+1}$ while increasing $p_{i,j}$ is feasible and will yield higher throughput.
\end{proof}

As we can see, the energy harvested in a single phase can support SN's data transmission in the next phase. Therefore, the optimization problem in Eq. (\ref{initial}) is simplified as Eq. (\ref{obj_fin_relay2}), and the solution will be discussed in Section IV-B.
\begin{equation}
\label{obj_fin_relay2}
\begin{array}{l}
\mathop {\max }\limits_{\{ {p_{2,j}},{\alpha_2}\} } \sum\limits_{j = 1}^N {\log (1 + {p_{2,j}}{\gamma _2})} \\
\begin{array}{*{20}{l}}
{s.t.}&{P_{1,1}}=P_{1,0}\text{, }{P_{1,2}}=({p_{2,1}+\alpha_2})\beta\\
{}&{P_{1,j}}=p_{2,j-1}\beta,j\in\{3,\cdots,N\}\\
AC_{2}:&{p_{2,2}}\le{(p_{2,1}+\alpha_2)}\beta\gamma\\
AC_{j}:&{p_{2,j}}\le{p_{2,j-1}}\beta\gamma,j\in\{3,\cdots,N\}\\
EC_{1,1}:&{p_{2,1}}\le{P_{1,0}}\gamma\\
EC_{2,1}:&{{p_{2,1}} \le P_{2,0}+{P_{1,0}}\beta-\alpha_2}\\
EC_{2,j}:&\sum\limits_{k = 1}^j {{p_{2,k}}} \le {P_{2,0}}+{P_{1,0}}\beta -\alpha_2(1-\beta^2)\\
&\quad\quad\quad\quad\quad\beta^2 \sum\limits_{k = 1}^{j-1} {{p_{2,k}}}
\end{array}
\end{array}
\end{equation}
where the additional constraints $AC_{2}$ and $AC_{j}$, $j\in\{3,\cdots,N\}$ are used to guarantee the feasibility of the fully greedy strategy of SN. $EC_{2,j}$, $j\in\{2,\cdots,N\}$ is the EC-constraint of RN with fully greedy SN in the $j$-th phase. $EC_{2,N}$ is satisfied with equality.

\subsection{$\beta\gamma<1$}
In this case, suppose $\beta{p_{1,j}}$ is directly used by RN to forward data in the $j$-th phase, then, instead of $p_{2,j}$, only $p_{2,j}(\gamma-\beta)/\gamma$ is required from RN's residual power. The scenario is equivalent to a system where: 1) channel condition of the RN-DN link is improved by the factor of $\gamma/(\gamma-\beta)$; 2) SN can harvest energy from both $p_{2,j}$ and $\alpha_{2,j}$; 3) RN can only harvest energy from $\alpha_{1,j}$. 

Consequently, 1) for the \emph{data transmission} part, the SNRs of the SN-RN link and RN-DN link are $\gamma'_1=\gamma_1$ and $\gamma'_2=\gamma_1/(\gamma-\beta)$, respectively. The SNR ratio is $\gamma'=\gamma-\beta$. The corresponding harvesting efficiency is $\beta'=\beta\gamma/\gamma'$. 2) For the \emph{power supplement} part, the harvesting efficiency is still $\beta$.

\emph{Note}: In this section, $p_{2,j}$ represents the transmission power of the equivalent system. For the original system, the transmission power of RN in the $j$-th phase is
\[P_{2,j}=p_{2,j}\gamma/\gamma'+\alpha_{2,j}\]

\emph{Proposition 3:} There exists an optimal power allocation which satisfies the following equality. 
\[p_{i,j}=p_{i,j+1}\ge p_{i,N},i\in\{1,2\}, j\in\{2,\cdots,N-2\}\] 
\begin{proof}
For $j\ge2$, the power supplement exists only in one specific direction, from SN to RN, i.e., $\alpha_{2,j}=0$. Then, similar to Eq. (\ref{useful}), it can be proven the solution with $p_{i,j}<p_{i,j+1}$ is not optimal. So, $p_{i,j}\ge p_{i,j+1},i\in\{1,2\}, j\in\{2,\cdots,N-1\}$.

We assume RN uses the initial power storage $P_{2,0}$ first, when it is depleted, RN then asks for power supplements from SN. The power supplements are partitioned such that RN has just enough power to forward the data received in each phase. As illustrated in Fig. \ref{onlyOne}, if $\alpha_{1,k}=0$, then $\alpha_{1,j}=0$, $j\in\{1,\cdots,k-1\}$. If $\alpha_{1,k}>0$, then $\alpha_{1,j}>0$, $j\in\{k+1,\cdots,N\}$.
\begin{figure}
\centering
\captionsetup[figure]{skip=0pt}
\includegraphics[width=3 in]{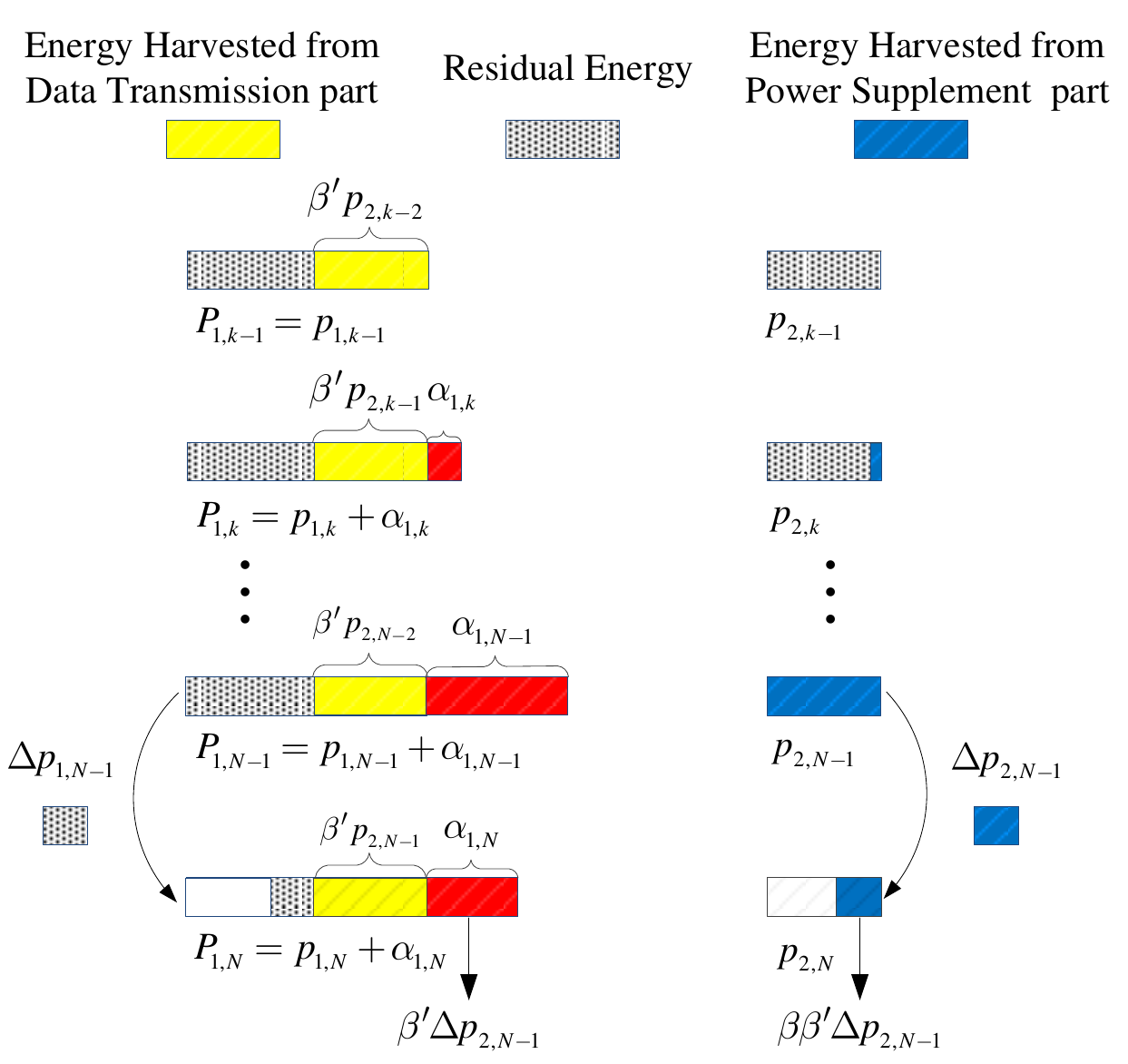}
\caption{Power Supplement Partition.}
\label{onlyOne}
\vspace{-1em}
\end{figure}

Suppose in the optimal solution $\exists j \in \{ 2, \cdots ,N-2\} $ with $p_{i,j}>{p_{i,j+1}}$, then, depending on $\alpha_{1,j+1}$ and $\overline {{P_{1,j}}}$, new solutions can always be found that will increase the aggregate throughput of the system.\\
a) $\alpha_{1,j+1}=0$\\
a.1) $\overline {{P_{1,j}}}\ge (p_{1,j}+p_{1,j+1})({{2 - \beta' \gamma' }})/2$:
\[\left\{ \begin{array}{l}
p{^*_{i,j}} = p{^*_{i,j+1}} = ({p_{i,j}} + p_{i,j+1})/2,i\in\{1,2\}\\
\overline{P^*_{i,j+2}}=\overline{P_{i,j+2}}
\end{array}\right.\]
where $\overline{P^*_{i,j+2}}$ is the residual power of SN ($i=1$) and RN ($i=2$) with new power allocation.\\
a.2) $\overline {{P_{1,j}}}< (p_{1,j}+p_{1,j+1})({{2 - \beta' \gamma' }})/2$:
\[\left\{ \begin{array}{l}
p{^*_{1,j}} = p{^*_{1,j+1}}= {{\overline {{P_{1,j}}}}}/({{2 - \beta' \gamma' }})\\
p{^*_{2,j}} = p{^*_{2,j+1}} = {{\overline {{P_{1,j}}} \gamma' }}/({{2 - \beta' \gamma' }})
\end{array}\right.\]
It can be checked that the following equation is satisfied. 
\[\mathop{\min }\limits_{i=1,2}\{\overline {P_{i,j+2}}\gamma'_i\}\le\mathop{\min }\limits_{i=1,2}\{\overline {P^*_{i,j+2}}\gamma'_i\} \le p{^*_{i,j+1}}\gamma'_i\]
From a), we can see that although the throughput of phase $j$ and $j+1$ may not increase, the aggregate throughput from phase $j$ to phase $N$ will increase, with $p_{i,j}=p_{i,j+1},i\in\{1,2\}, j\in\{2,\cdots,N-2\}$.\\
b) $\alpha_{1,j+1}>0$: As we can see in Fig. \ref{onlyOne}, for $j=N-1$, if $p_{2,N-1}>p_{2,N}$, decreasing $p_{i,N-1}$ and increasing $p_{i,N}$ until $p^*_{2,N-1}=p^*_{2,N}$ or $\alpha^*_{1,N}=0$ will yield higher throughput. In fact, for any $j>2$, if $\alpha_{1,j+1}>0$, $p_{i,j}=p_{i,j+1}$.
%
\end{proof}
\emph{Remark 2:} If SN provides power supplements, $p_{i,j}=p_{i,j+1},i\in\{1,2\},j\in\{2,\cdots,N-1\}$. Furthermore, if $\alpha_{2}=0$, $p_{i,j}=p_{i,j+1},i\in\{1,2\}, j\in\{1,\cdots,N-1\}$.

\emph{Proposition 4:} With $\beta\gamma<1$, $\beta<\gamma$, any power supplements can be aggregated to the first phase.
\begin{equation}
\label{bothless1}
\alpha_1\alpha_2=\alpha_{i,j}=0,i\in\{1,2\},j\in\{2,\cdots,N\}
\end{equation}
where $\alpha_i\ge0$ is the aggregated \emph{power supplement} transmitted by SN ($i=1$) and RN ($i=2$) in the first phase.
\begin{proof}
Let $k$ be the first phase where SN provides positive power supplement. a) $k\in\{1,2\}$: $\alpha_2\alpha_{1,k}>0$ is not optimal (\emph{Remark 1}). b) $k>2$: Since $p_{i,j}=p_{i,j+1}$ , $j\in\{2,\cdots,k-1\}$ (\emph{Proposition 3}) $\beta'p_{2,j}=\beta'\gamma'p_{1,j}<p_{1,j+1}$. Then, the power supplement $\alpha_{1,k}$ must come from the residual power of SN at the beginning of phase 2. Thus, it can be aggregated as follows:
\begin{equation}\label{itagian}
\alpha^*_{1,2}=\alpha_{1,k}\text{, }\alpha^*_{1,k}=0
\end{equation}
Since $\alpha_2\alpha^*_{1,2}>0$ is not optimal, we have $\alpha_2\alpha_{1,j}=0$, $j\in\{1,\cdots,N\}$. 

Furthermore, if SN provides power supplements, $p_{1,j}=p_{1,j+1}$, $j\in\{1,\cdots,N-1\}$ (\emph{Remark 2}). Similar to Eq. (\ref{itagian}), all of the power supplements can be aggregated to the first phase, i.e., $\alpha_1$. 
\end{proof}

For the equivalent system where RN can only harvest energy from $\alpha_1$, the energy causality constraints of RN, i.e., $EC_{2,j}$, $j\in\{1,\cdots,N\}$, becomes the constant power budget. Applying \emph{Proposition 4}, the optimization problem in Eq. (\ref{initial}) is simplified as follows:
\begin{equation}
\label{obj_fin_relay3}
\begin{array}{l}
\mathop {\max }\limits_{\{ {p_{2,j}},{\alpha_1},{\alpha_2}\} } \sum\limits_{j = 1}^N {\log (1 + {p_{2,j}}{\gamma' _2})} \\
\begin{array}{*{20}{l}}
{s.t.}&\alpha_1\alpha_2=0\\
EC_{2,N}:&{\sum\limits_{j = 1}^N {{p_{2,j}}}  + {\alpha_2} = {P_{2,0}}+{\alpha_1}\beta}\\
EC_{1,1}:&{{p_{2,1}} \le ({P_{1,0}}-\alpha_1)\gamma'}\\
EC_{1,j}:&{\sum\limits_{k = 1}^j {{p_{2,k}}}  \le ({P_{1,0}}-\alpha_1 + \beta{\alpha_2})\gamma'}+ \\
&\quad\quad\quad\quad{\beta' \gamma'\sum\limits_{k = 1}^{j - 1} {{p_{2,k}}} }\\
\end{array}
\end{array}
\end{equation}
where $EC_{2,N}$ is the constant power budget of the equivalent system. $EC_{1,j}$, $j\in\{2,\cdots,N\}$ is the EC-constraint of SN with aggregated power supplements.

Since $\alpha_1>0$ indicates $\alpha_2=0$ and equal power allocation of RN (\emph{Remark 2}), with $\beta' \gamma'=\beta\gamma<1$, $EC_{1,j}$, $j\in\{1,\cdots,N\}$ can be simplified as follows:
\[{P_{2,0}+{\alpha_1}\beta}\le(P_{1,0}-\alpha_1)\gamma'+\beta' \gamma'(N-1)\frac{{P_{2,0}+{\alpha_1}\beta}}{N}\]
So, $\alpha_1>0$ requires ${P_{1,0}} \ge {P_{2,0}}\frac{{N - (N - 1)\beta' \gamma' }}{{N\gamma' }}$. 

Depending on the value of $P_{1,0}$, the solutions to Eq. (\ref{obj_fin_relay3}) are given in Table \ref{tab2}. For case 2, $\alpha^*_2$ can be derived using the Lagrange method, which is not shown here because of the space limit.

\captionsetup[table]{skip=0pt}
\begin{table}[ht]
\caption{Optimal Solution with $\beta<\gamma$, $\beta\gamma<1$} 
\label{tab2} 
\centering\  
\begin{tabular}{|L|}
\hline                        
Case 1: ${P_{1,0}} \ge {P_{2,0}}\frac{{N - (N - 1)\beta' \gamma' }}{{N\gamma' }}$\\\hline
$\left\{\begin{array}{l}
\alpha^*_2=0\\
P_{1,0}-\alpha^*_1=({P_{2,0}+{\alpha^*_1}\beta})\frac{N-(N-1)\beta'\gamma'}{N\gamma'}\\
{p^*_{2,j}}=({P_{2,0}+{\alpha^*_1}\beta})/N,j\in\{1,\cdots,N\}
\end{array}\right.$
\\\hline 
Case 2: ${P_{1,0}} < {P_{2,0}}\frac{{N - (N - 1)\beta' \gamma' }}{{N\gamma' }}$\\\hline
$\left\{\begin{array}{l}
\alpha^*_1=0\text{, }{p^*_{2,j}}={p_{2,j+1}},j\in\{2,\cdots,N-2\}\\
{p^*_{2,1}},{p^*_{2,N}}\le{p^*_{2,j}},j\in\{2,\cdots,N-1\}\\
{p^*_{2,1}}+\alpha^*_2\ge{p^*_{2,j}},j\in\{2,\cdots,N-1\}\\
{p^*_{2,1}}\in\{P_{1,0}\gamma',p^*_{2,2}\}$, ${p^*_{2,N}}\in\{\overline {P_{2,N}},p^*_{2,N-1}\}
\end{array}\right.$
\\\hline 
\end{tabular}
\vspace{-1em}
\end{table}

\emph{Remark 3:} For the scenarios with $\beta\ge\gamma$ and $\beta<\gamma$, $\beta\gamma\ge1$, the solutions should have the same structure as Eq. (\ref{obj_fin_relay3}) because ${\beta}^2$ in the EC-constraints of Eqs. (\ref{obj_fin_relay}) and (\ref{obj_fin_relay2}) is less than 1. The specific solutions are omitted due to space limit.
\section{Numerical Results}
\begin{figure*}
        \centering
        \begin{subfigure}[b]{2.3 in}
                \includegraphics[width=\textwidth]{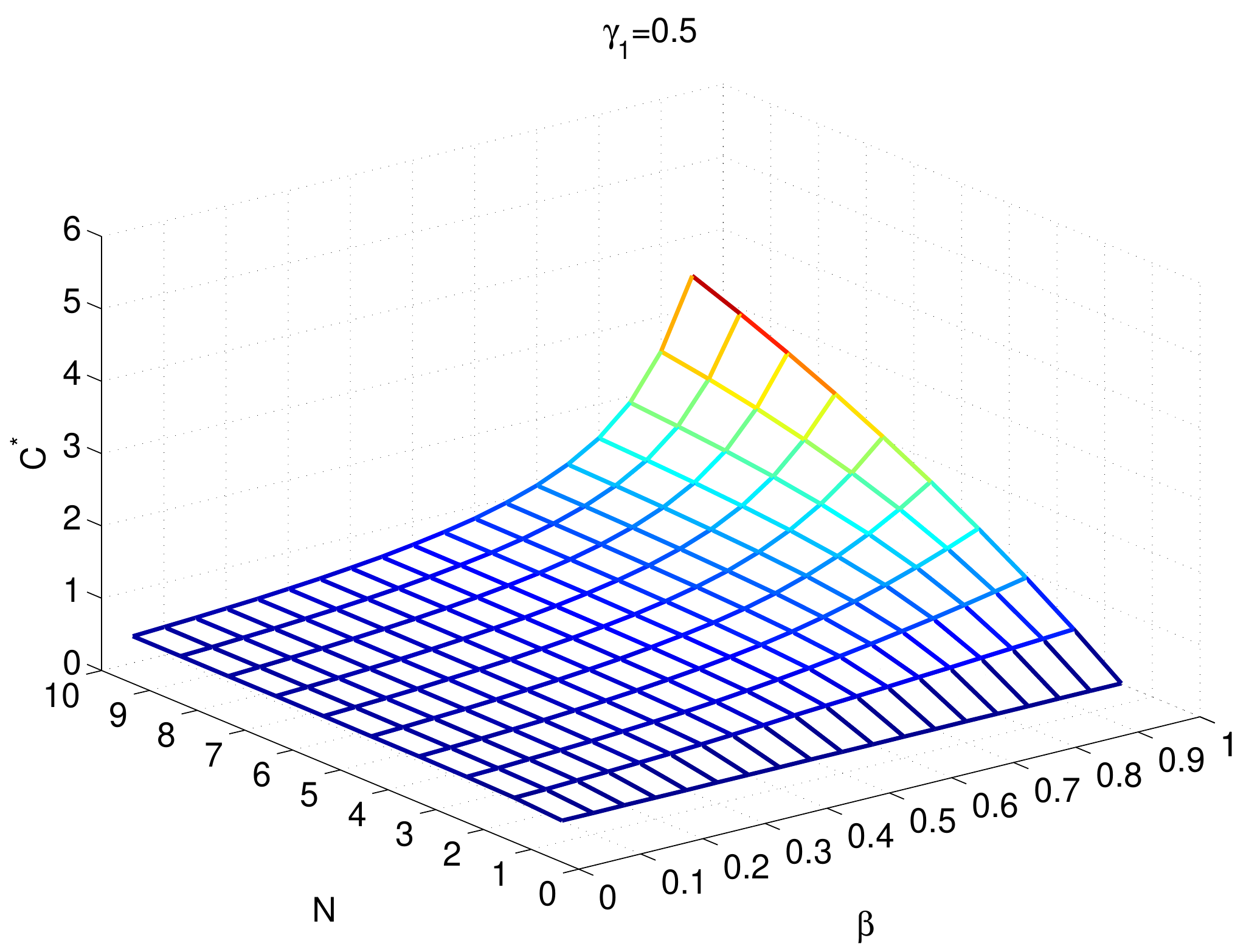}
        \end{subfigure}%
        ~ 
        \begin{subfigure}[b]{2.3 in}
                \includegraphics[width=\textwidth]{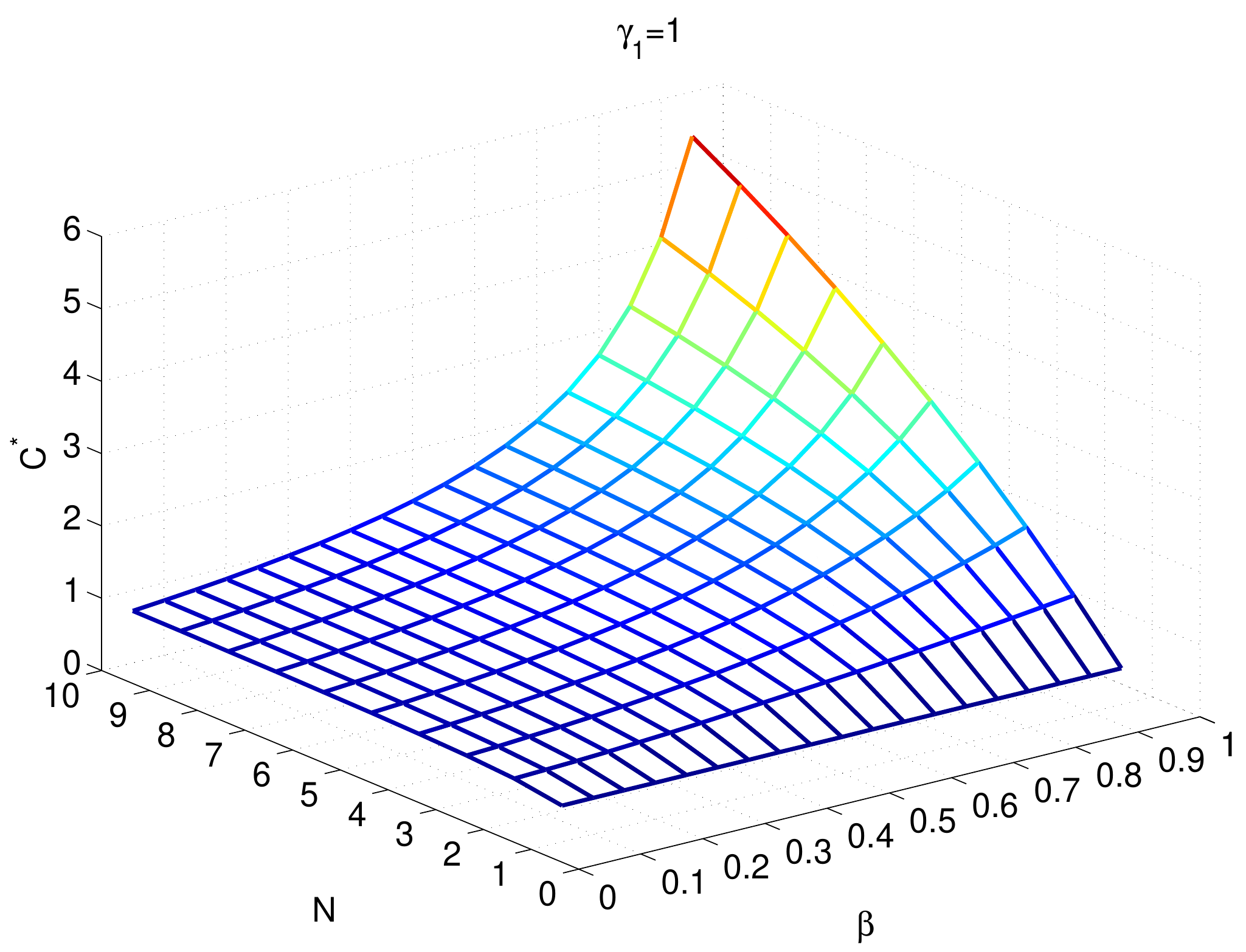}
        \end{subfigure}\\
        \begin{subfigure}[b]{2.3 in}
                \includegraphics[width=\textwidth]{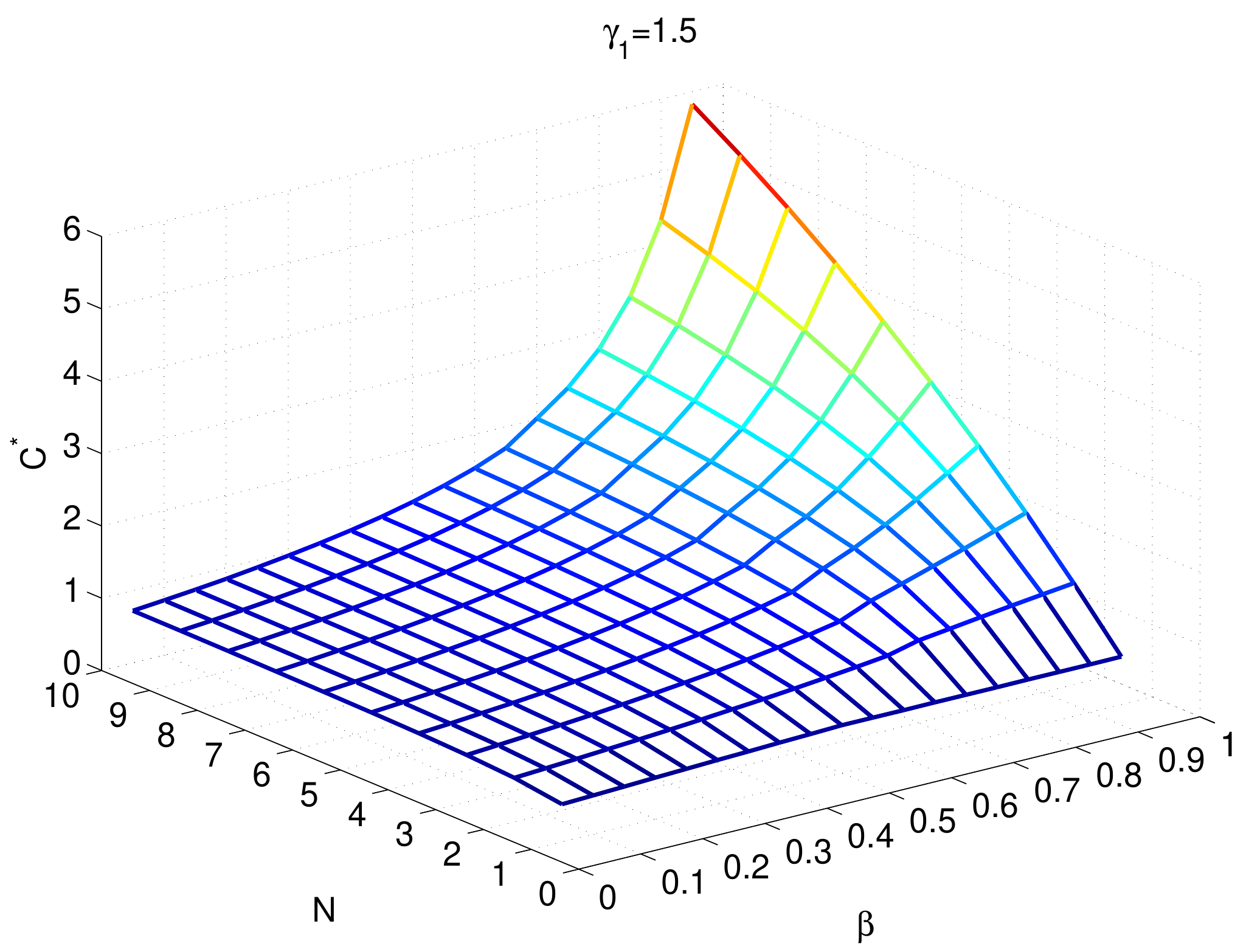}
        \end{subfigure}
        ~
          \begin{subfigure}[b]{2.3 in}
                \includegraphics[width=\textwidth]{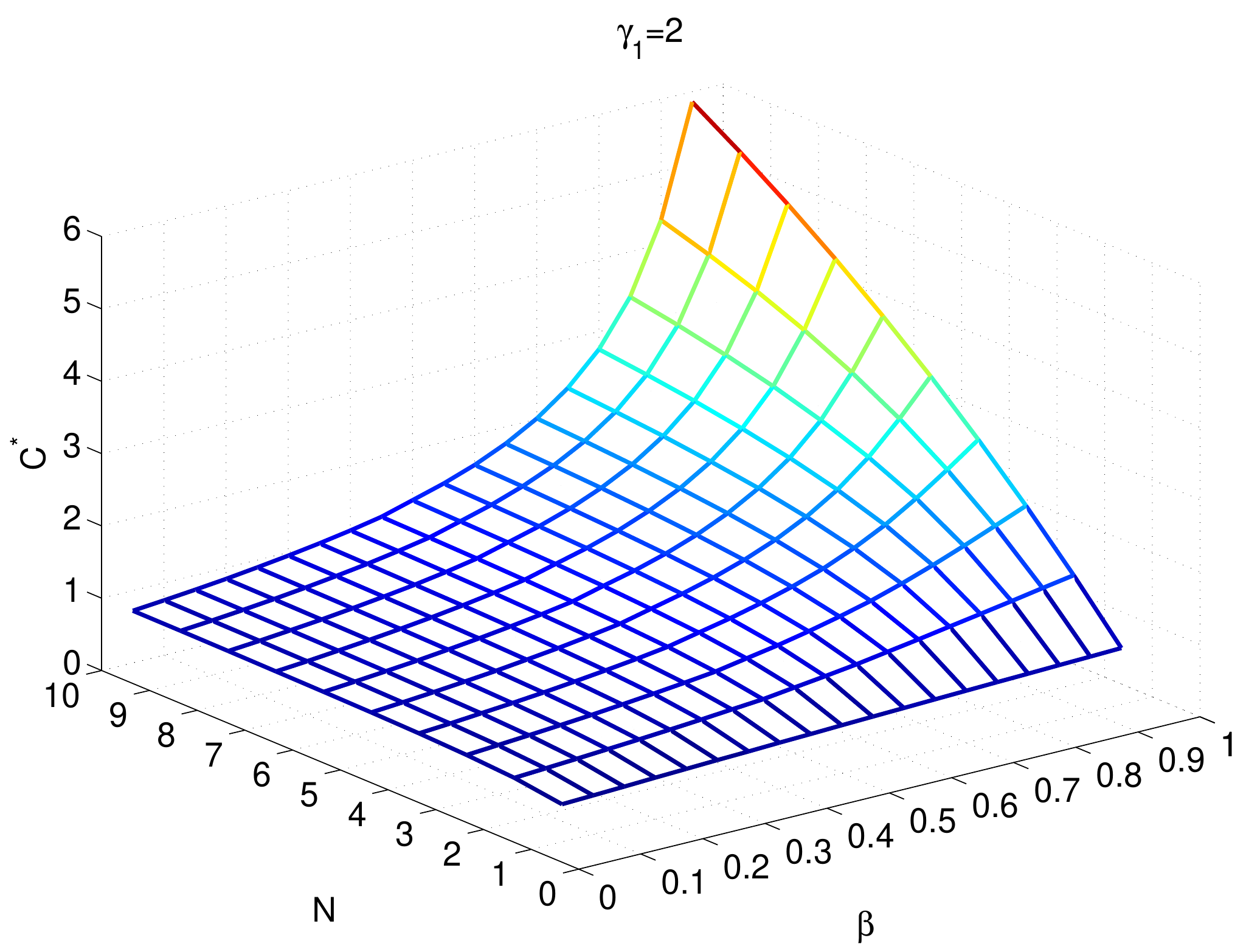}
        \end{subfigure}
        \caption{Optimal throughput vs. $\beta$, $N$ and $\gamma$ ($B=1$, $\gamma_2=1$, $P_{i,0}=1$, $i\in\{1,2\}$).}\label{fig_result}
        \vspace{-1em}
\end{figure*}

With unit bandwidth, unit initial power storage, and normalized SNR for RN-DN link, Fig. \ref{fig_result} provides the numerical results of the optimal power allocation (OPT) given in Table \ref{tab2}. As expected, the system throughput will increase with $N$ and harvesting efficiency $\beta$, and the performance improvement will be less obvious as SNR of SN-RN link increases. The reason is although high $\gamma_1$ can save SN's energy consumption, RN's energy consumption will determine the system throughput, as shown in Eq. (\ref{obj_fin_relay2}).

\begin{figure*}
        \centering
        \begin{subfigure}[b]{2.5 in}
                \includegraphics[width=\textwidth]{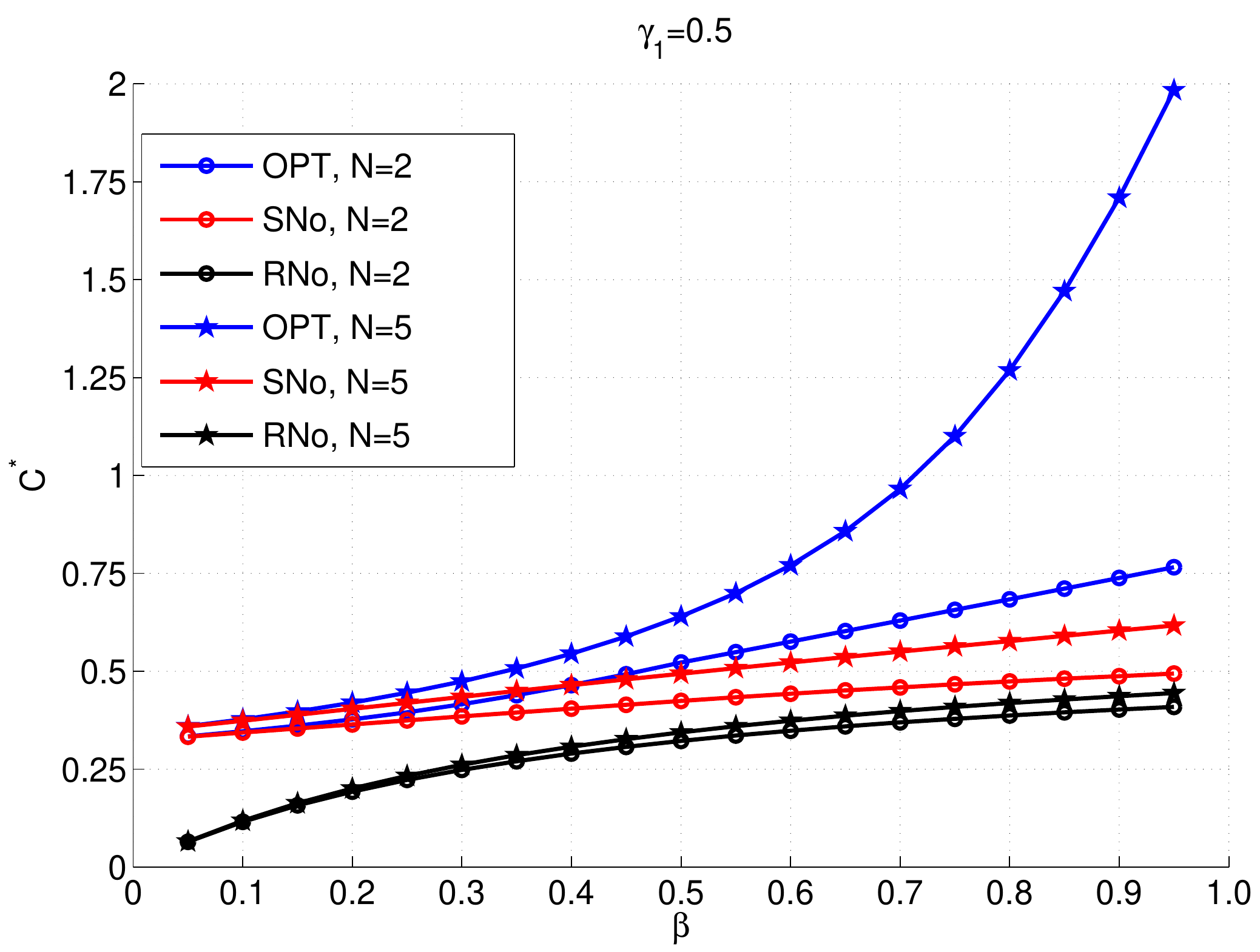}
        \end{subfigure}
        ~
          \begin{subfigure}[b]{2.5 in}
                \includegraphics[width=\textwidth]{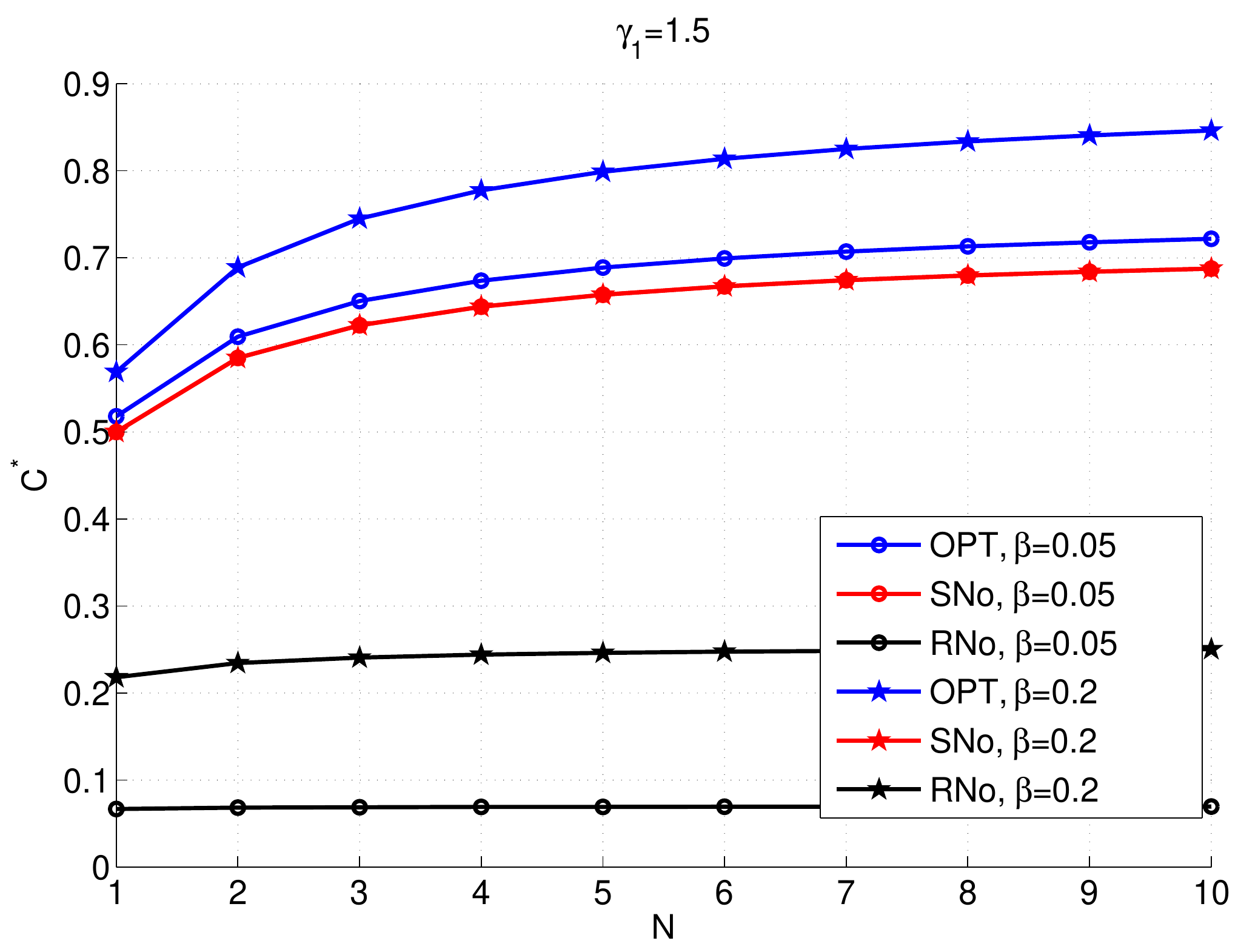}
        \end{subfigure}
        \caption{Throughput comparisions with SNo and RNo algorithms.}\label{fig:result}
        \vspace{-1em}
\end{figure*}
The SN-only (SNo) and RN-only (RNo) power allocation algorithms are used to provide performance reference for the OPT power allocation. The SNo algorithm is designed for a system with EH-SN and regular RN, which can only rely on its own initial power storage $P_{2,0}$. The RNo algorithm is designed for a system where the regular SN has total power supply of $P_{1,0}+P_{2,0}$, and EH-RN uses the harvested energy to forward data.

As illustrated in Fig. \ref{fig:result}, the performance difference between the OPT and SNo algorithms benefits from the simultaneous energy and energy reception of EH-RN, while the performance difference between SNo and RNo algorithms indicates that in the half duplex relay system, the harvest energy by EH-SN in the even TSs can be used to improve the throughput. For the SNo and RNo algorithms, the increment in the overall throughput is less obvious as $N$ and $\beta$ increase to a certain point where the power resource of the relay node is more stringent.
\section{Conclusion}
To study the wireless energy sharing and simultaneous data and energy transfer, we have designed the joint energy management policies for the RF-EH enabled relay system. It can be seen that RF energy harvesting can benefit the energy aware wireless communications by improving energy harvesting efficiency, utilizing more time to transmit delay tolerant traffic, and/or sharing the energy when the networks have unbalanced energy or traffic distribution.
\bibliographystyle{IEEEtran}
\bibliography{mybib}
\end{document}